\documentclass[aps,floatfix,showpacs,superscriptaddress,showkeys,twocolumn]{revtex4} 

\usepackage{amssymb,amsmath,amsfonts,latexsym,graphicx,epsfig,here}

\usepackage{mkPhys}

\newcommand{\beaa}{\begin{eqnarray*}} 
\newcommand{\enaa}{\end{eqnarray*}}

\newcommand{\bea}{\begin{eqnarray}}
\newcommand{\ena}{\end{eqnarray}} 

\newcommand{\be}{\begin{equation}}
\newcommand{\en}{\end{equation}}

\newcommand{\nn}{\nonumber\\}

\begin{document}

\title{Quark model description of the tetraquark state X(3872) in a 
\\relativistic constituent quark model  
with infrared confinement}


\author{S. Dubnicka}
\affiliation{Institute of Physics
Slovak Academy of Sciences
Dubravska cesta 9
SK-842 28 Bratislava, Slovak Republic}
\author{A. Z. Dubnickova}
\affiliation{Comenius University
Dept. of Theoretical Physics
Mlynska Dolina
SK-84848 Bratislava, Slovak Republic }
\author{M. A. Ivanov}
\affiliation{Bogoliubov Laboratory of Theoretical Physics, 
Joint Institute for Nuclear Research, 141980 Dubna, Russia}
\author{J. G. K\"{o}rner}
\affiliation{Institut f\"{u}r Physik, Johannes Gutenberg-Universit\"{a}t,
D--55099 Mainz, Germany}


\begin{abstract} 
We explore the consequences of treating the X(3872) meson as a tetraquark
bound state. As dynamical framework we employ a relativistic constituent 
quark model which includes infrared confinement in an effective way. 
We calculate the decay widths of the observed
channels $X\to \Jpsi+2\pi (3\pi)$ and $X\to \bar D^0 +  D^0+\pi^0$ via the
intermediate off--shell states $X\to \Jpsi+\rho(\omega)$ and 
$X\to \bar D +  D^\ast$.
For reasonable values of the size parameter $\Lambda_X$ of the 
X(3872) we find consistency with the available experimental data.
We also discuss the possible impact of the X(3872) in a s-channel dominance
description of the $\Jpsi$ dissociation cross section.

\end{abstract}



\maketitle

\section{Introduction}
\label{sec:intro}

A narrow charmonium--like state $X(3872)$ was observed in 2003 
in the exclusive decay process $B^\pm\to K^\pm\pi^+\pi^-\Jpsi$
\cite{Choi:2003ue}. 
The  $X(3872)$ decays into $\pi^+\pi^-\Jpsi$ and has a mass of 
$m_X=3872.0 \pm 0.6 ({\rm stat}) \pm 0.5 ({\rm syst}) $
very close to the $M_{D^0}+M_{D^{\ast\,0}}=3871.81 \pm 0.25$ mass threshold 
\cite{PDG}.
Its width was found to be less than 2.3 MeV at $90\%$ confidence level.
The state was confirmed in B-decays by the \babar experiment 
\cite{Aubert:2004fc} 
and in $p\overline{p}$ production
by the Tevatron experiments CDF \cite{Acosta:2003zx} and D\O{}
\cite{Abazov:2004kp}. The most precise measurement up to now was done
in \cite{CDF} with $M_X=3871.61 \pm 0.16 \pm 0.19 $. The new average
mass given in \cite{Acosta:2003zx} is
\be
M_X=3871.51 \pm 0.22 \, {\rm MeV}.
\en  

From the  observation of the decay $X(3872)\rightarrow \Jpsi \gamma$ reported 
by both the Belle \cite{jpsigamma_Belle} and \babar  \cite{jpsigamma_BaBar} 
collaborations and from the angular analysis performed by CDF 
\cite{Abulencia:2006ma},  it was shown that the only quantum numbers 
compatible with the data are $J^{PC}=1^{++}$ or $2^{-+}$. 
However, the observation of the decays into 
$D^0\overline{D}^{0}\pi^0$ by the Belle and \babar collaborations 
\cite{Gokhroo:2006bt,Aubert:2007rva} allows one to exclude the choice $2^{-+}$
because the near-threshold decay 
$X\to D^0\overline{D}^{0}\pi^0$   is expected to be strongly suppressed 
for $J=2$.

The Belle collaboration has  reported evidence for the decay mode  
$X \to \pi^+\pi^-\pi^0 \Jpsi$ with  a strong three-pion peak between 750 MeV
and the kinematic limit of 775 MeV \cite{jpsigamma_Belle}, suggesting that 
the process 
is dominated by the  sub-threshold decay $X \to \omega \Jpsi$. 
It was found that the branching ratio
of this mode is almost the same as that of the mode $X \to \pi^+\pi^- \Jpsi$:
\be
\frac{ {\cal B}(X\to\Jpsi\pi^+\pi^-\pi^0) }
     { {\cal B}(X\to\Jpsi\pi^+\pi^-) }
 = 1.0 \pm 0.4 ({\rm stat}) \pm 0.3 ( {\rm syst} ).
\label{eq:expt}
\en
These observations imply strong isospin violation because the three-pion decay
proceeds via an intermediate $\omega$-meson with isospin 0 whereas
the two-pion decay proceeds via the intermediate $\rho$-meson with isospin 1.
Also the two-pion decay via the intermediate $\rho$-meson is very difficult
to explain by using an interpretation of the $X(3872)$ as a
simple $c\bar c$ charmonium state with isospin 0.  There are several
different interpretations of the $X(3872)$ in the literature:
\begin{itemize}
\item The $X(3872)$ is a molecule bound state ($D^0\overline{D}^{\ast\,0}$) 
with small binding energy 
     (see, Refs.~(\cite{Tornqvist:2004qy}-\cite{FernandezCarames:2010de}));
\item The $X(3872)$ is a tetraquark state composed of a diquark and antidiquark
     (see, Refs.~(\cite{Maiani:2004vq}-\cite{Terasaki:2009in}));
\item  The $X(3872)$ should be interpreted in terms of threshold cusps 
\cite{Bugg:2004rk};
\item The $X(3872)$ should be interpreted in terms of hybrids 
\cite{Li:2004sta} and glueballs \cite{Seth:2004zb}.
\end{itemize} 
A description of the current theoretical and experimental situation for the 
new charmonium states may be found in the reviews
\cite{Eichten:2007qx}-\cite{Abe:2004zs}. 


The most intriguing question at present is whether the X(3872) is a 
losely-bound charm-meson molecule with a binding energy of
$M_X-(M_{D^{\ast\,0}}+M_{D^0}) = -0.30\pm 0.40$ MeV,
or a tetraquark composed of a color diquark and color antidiquark.
Recently, the molecular nature of the the X(3872) has been carefully 
discussed in the literature.
The authors of~\cite{Bignamini:2009sk} assumed that
the X(3872) is a $D^0\bar D^{\ast\,0}$ molecule and, based on this assumption,
estimated its
prompt production cross section at the Fermilab Tevatron. 
They presented a theoretical upper limit
on the prompt production cross section which is about 30 times smaller 
than the observed prompt production rate at the Tevatron.   
The conclusion was that S-wave resonant scattering is unlikely to
allow the formation of a loosely bound $D^0\bar D^{\ast\,0}$ molecule
in high energy hadron collisions.
In reference~\cite{Artoisenet:2009wk} it was argued that one
needs to take into account charm-meson rescattering in the 
analysis of~\cite{Bignamini:2009sk}. As a consequence the theoretical upper 
limit is increased by orders of magnitude
and is then compatible with the observed production rate at the Tevatron.
This conclusion was later on criticized in 
\cite{Bignamini:2009fn}. First, using the results of~\cite{Artoisenet:2009wk},
the authors of~\cite{Bignamini:2009fn} argue that a new unobserved
$X_s(4080)$ molecule composed of a $D_s\bar D_s^\ast$ pair
should have been observed at the Tevatron. Second, they cast some doubts
on the applicability of the Watson theorem for final state interactions
in the above calculation.


The tetraquark state interpretation of the X(3872) 
was successfully applied to describe the available experimental data
on the decays $X\to\Jpsi\pi^+\pi^-$ and $X\to\Jpsi\pi^+\pi^-\pi^0$ 
\cite{Maiani:2004vq}.
By using an effective three-meson Lagrangian with the coupling
taken from a similar analysis of light scalar mesons,
the authors found the width of the X-meson to be 1.6 MeV in accordance
with the experimental bound. Contrary to this, the values of the widths
calculated by using QCD sum rules were found 
to be too large - around 50 MeV \cite{Nielsen:2009uh}.

In this paper we provide an independent analysis of the the properties of the 
X(3872) meson which we interpret as a tetraquark state as in 
\cite{Maiani:2004vq}.
We work in the framework of the relativistic constituent quark model which 
has recently been extended to include infrared confinement effects 
\cite{Branz:2009cd}. 
The improved model \cite{Branz:2009cd} is a successful generalization of the 
relativistic constituent quark 
model which some of us have developed over many 
years~\cite{light_mes}-\cite{Efimov:1993ei}.
The relativistic constituent quark model can be viewed as an effective 
quantum field theory approach to hadronic interactions 
based on an interaction Lagrangian of hadrons interacting with their  
constituent quarks. Once the relevant interpolating quark current is written
down one can evaluate the matrix elements of the physical processes
in a self-consistent way. The nice feature of this approach is that
multiquark systems as e.g. baryons and tetraquarks 
can be treated on the same footing as the simplest quark-antiquark states.
The coupling strength of hadrons with their interpolating quark currents 
is determined by  the compositeness condition~$Z_H=0$  \cite{SWH}
where $Z_H$ is the wave function renormalization constant of the hadron.
Matrix elements are generated by a set of quark loop diagrams 
according to a $1/N_c$ expansion. The ultraviolet divergences of the quark 
loops are regularized by including vertex form factors for the hadron-quark 
vertices which, in addition, describe finite size effects due to the 
non-pointlike structure of hadrons. 
The relativistic constituent quark model contains only a few model parameters:
the light and heavy constituent quark masses, 
the confinement scale, and the size 
parameters that describe the size of the distribution of the constituent 
quarks inside the hadron.
 
The paper is organized as follows. In Sec.~II we construct the nonlocal
generalization of the four-quark interpolating current 
of the X(3872) written down in \cite{Maiani:2004vq}. This leads to a
nonlocal effective Lagrangian describing the interaction
of the X(3872) meson with its constituent quarks. The coupling strength of
the X(3872) w.r.t. its constituent quarks is 
determined from the compositeness condition $Z_{H}=0$.  
We also briefly discuss the implementation of infrared confinement
in our scheme.
In Sec.~III we calculate  the matrix elements of the transitions
$X\to \Jpsi +\rho(\omega)$ and $X\to D + \bar D^\ast$.
The results are then used to evaluate the widths of the decays
$X\to \Jpsi+2\pi(3\pi)$ and $X\to D^0+\bar D^0 + \pi^0$.
In Sec.~IV we present the results of our numerical analysis and compare our 
results with the 
results of other approaches. Finally, in Sec.~V we summarize our findings.

\section{Theoretical framework}

\subsection{Effective Lagrangians}

The authors of \cite{Maiani:2004vq} suggested to consider the $X(3872)$ meson
as a $J^{PC}=1^{++}$ tetraquark state with a symmetric spin distribution:
$[cq]_{S=0}\,[\bar c \bar q]_{S=1} + [cq]_{S=1}\,[\bar c \bar q]_{S=0}$,
$(q=u,d)$. The nonlocal version of the four-quark interpolating current
reads
\begin{widetext}
\bea
J^\mu_{X_q}(x) &=& \int\! dx_1\ldots \int\! dx_4 
\delta\left(x-\sum\limits_{i=1}^4 w_i x_i\right) 
\Phi_X\Big(\sum\limits_{i<j} (x_i-x_j)^2 \Big)
\label{eq:cur}\\
&\times&
\tfrac{1}{\sqrt{2}}\, \varepsilon_{abc}\varepsilon_{dec} \,
\Big\{\, [q_a(x_4)C\gamma^5 c_b(x_1)][\bar q_d(x_3)\gamma^\mu C \bar c_e(x_2)]
      +[q_a(x_4)C\gamma^\mu c_b(x_1)][\bar q_d(x_3)\gamma^5 C \bar c_e(x_2)]
\,\Big\},\nn
&&\nn
w_1&=&w_2 = w_c =\frac{m_c}{2(m_q+m_c)}, \qquad
w_3 = w_4 = w_q =\frac{m_q}{2(m_q+m_c)}. \nonumber
\ena 
\end{widetext}

The matrix $C=\gamma^0\gamma^2$ is the charge conjugation 
matrix: $C=C^\dagger=C^{-1}=-C^T$, $C\Gamma^TC^{-1}=\pm\Gamma$, ($"+"$ for
$\Gamma=S,P,A$ and $"-"$ for $\Gamma=V,T$). The numbering of the coordinates
$x_i$ is chosen such that one has a convenient arrangement of
vertices and propagators in the Feynman diagrams.
The effective interaction Lagrangian describing the coupling
of the meson $X_q$ to its constituent quarks is written in the form
\be
{\cal L}_{\rm int} = g_X\,X_{q\,\mu}(x)\cdot J^\mu_{X_q}(x), \qquad (q=u,d).
\label{eq:lag}
\en     
The state $X_u$ breaks isospin symmetry maximally:
\be
X_u = \frac{1}{\sqrt{2}}\Big\{  
            \underbrace{ \frac{X_u+X_d}{\sqrt{2}} }_{I=0}
          + \underbrace{ \frac{X_u-X_d}{\sqrt{2}} }_{I=1}
                       \Big\}.
\en
The authors of \cite{Maiani:2004vq} take the physical states to be a linear
superposition of the $X_u$ and $X_d$ states 
according to
\bea  
X_l\equiv X_{\rm low} &=&\hspace{0.2cm}  X_u\, \cos\theta +  X_d\, \sin\theta,\nn
X_h\equiv X_{\rm high} &=& - X_u\, \sin\theta +  X_d\, \cos\theta.
\label{eq:mixing}
\ena
The mixing angle $\theta$ can be determined from fitting the ratio
of branching ratios Eq.~(\ref{eq:expt}).

The coupling constant $g_X$ in Eq.~(\ref{eq:lag}) will be determined from
the compositeness condition $Z_{H}=0$, see, e.g. Refs.~\cite{SWH},
\cite{Efimov:1993ei}, \cite{Branz:2009cd}. The compositeness condition
requires that the renormalization constant $Z_X$ of the 
elementary meson $X$ is set to zero, i.e.
\be
\label{eq:Z=0}
Z_X = 1-\Pi_X^\prime(m^2_X)=0,
\en
where $\Pi_X(p^2)$ is the scalar part of the vector-meson mass operator
\bea
\Pi^{\mu\nu}_X(p) &=& g^{\mu\nu} \Pi_X(p^2) + p^\mu p^\nu \Pi^{(1)}_X(p^2),\nn
 \Pi_X(p^2) &=& \frac{1}{3}\left(g_{\mu\nu}-\frac{p_\mu p_\nu}{p^2}\right)
                \Pi^{\mu\nu}_X(p).
\label{eq:mass}
\ena

The Fourier transform of the vertex  function
$\Phi_X\Bigl(\sum\limits_{i<j} ( x_i - x_j )^2 \Bigr)$ can be calculated
by using appropiately chosen Jacobi coordinates. One has
\bea
x_1&=&x  +  \frac{2w_2+w_3+w_4}{2\sqrt{2}} \rho_1 
           -  \frac{w_3-w_4}{2\sqrt{2}} \rho_2 
           +  \frac{w_3+w_4}{2}\rho_3, \nn
x_2&=&x  -  \frac{2w_1+w_3+w_4}{2\sqrt{2}} \rho_1 
           -  \frac{w_3-w_4}{2\sqrt{2}} \rho_2 
           +  \frac{w_3+w_4}{2}\rho_3, \nn
x_3&=&x  -  \frac{w_1-w_2}{2\sqrt{2}} \rho_1 
           +  \frac{w_1+w_2+2w_4}{2\sqrt{2}} \rho_2 
           -  \frac{w_1+w_2}{2}\rho_3, \nn
x_4&=&x  -  \frac{w_1-w_2}{2\sqrt{2}} \rho_1 
           -  \frac{w_1+w_2+2w_3}{2\sqrt{2}} \rho_2 
           -  \frac{w_1+w_2}{2}\rho_3, 
\nonumber
\ena 
where $x=\sum\limits_{i=1}^4 x_i w_i$ and 
$\sum\limits_{1\le i< j\le 4} (x_i-x_j)^2 =\sum\limits_{i=1}^3 \rho_i^2.$ 
One then has
\bea
&&
\Phi_X\Big(\sum\limits_{i<j} (x_i-x_j)^2 \Big) =
\prod_{i=1}^4 \int\!\!\frac{dp_i}{(2\pi)^4}\, 
e^{-i\sum\limits_{i=1}^4 p_i x_i}\nn
&&\hspace*{3cm}\times\,
\tilde \Phi_X(p_1,\ldots,p_4),\nn
&&\nn
&&
 \tilde \Phi_X(p_1,\ldots,p_4) = 
(2\pi)^4\,\delta\left({\textstyle\sum\limits_{i=1}^4 }p_i\right)\,
\bar \Phi_X(-\,\Omega^2),
\nn
&&
\bar \Phi_X(-\,\Omega^2) =
\tfrac{1}{4}\prod_{i=1}^3 \int\!\! d\rho_i\,
e^{i\,\sum\limits_{i=1}^3 \omega_i \rho_i}\, \Phi_X(R^2),
\label{eq:Fourier}
\ena
where $\Omega^2=\sum\limits_{i=1}^3 \omega_i^2$ and
      $R^2=\sum\limits_{i=1}^3 \rho_i^2 $.  
The Jacobi coordinates in momentum space read 
\bea
\omega_1 &=& \frac{p_1-p_2}{2\,\sqrt{2}}, \qquad
\omega_2 = \frac{p_1+p_2+2p_3}{2\,\sqrt{2}}, \nn
\omega_3 &=& \frac{p_1+p_2}{2}.
\label{eq:omega}
\ena
For calculational convenience we will choose a simple Gaussian form for the 
vertex function $\bar \Phi_X(-\,\Omega^2)$. 
The minus sign in the argument of this function is chosen to emphasize 
that we are working in Minkowski space. One has
\be
\bar \Phi_X(-\,\Omega^2) 
= \exp\left(\Omega^2/\Lambda_X^2\right)
\label{eq:Gauss}
\en
where the parameter $\Lambda_X$ characterizes the size of the X-meson.
Since $\Omega^2$ turns into 
$-\,\Omega^2$ in Euclidean space the form
(\ref{eq:Gauss}) has the appropriate fall-off behavior in the Euclidean region.
We emphasize that any choice for  $\Phi_X$ is appropriate
as long as it falls off sufficiently fast in the ultraviolet region of
Euclidean space to render the corresponding Feynman diagrams ultraviolet 
finite. As mentioned before we shall choose a Gaussian form for $\Phi_X$
for calculational convenience.

\begin{figure}[htbp]
\includegraphics[scale=0.4]{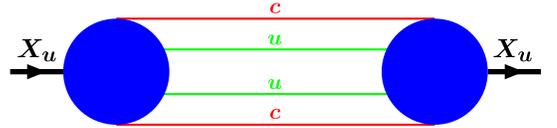} 
\caption{Diagram describing the $X_{u}$-meson mass operator.}
\label{fig:mass}
\end{figure}

We are now in the position to write down an explicit expression for
the derivative of the mass operator appearing in Eq.~(\ref{eq:Z=0}).
The corresponding three-loop diagram describing the X-meson mass operator is 
shown in 
Fig.~\ref{fig:mass}.

One has
%
\bea
&&
\Pi_X^\prime(p^2) = 
\frac{1}{2p^2}\,p^\alpha \frac{\partial}{\partial p^\alpha}\Pi_X(p^2) 
\nn
&&\nn
&&
=\frac{2\,g_X^2}{3\,p^2}\left(g_{\mu\nu}-\frac{p_\mu p_\nu}{p^2}\right)
\prod\limits_{i=1}^3\int\!\!\frac{d^4k_i}{(2\pi)^4i}\,
\bar\Phi_X^2\left(-\,K^2\right) 
\nn
&&
\times \Big\{
-w_c {\rm tr}
\left[S_c^{[12]}\!\not\! p S_c^{[12]}\gamma^5 S_q^{[2]} \gamma^5 \right]
{\rm tr}\left[S_c^{[3]}\gamma^\mu S_q^{[13]}\gamma^\nu \right] 
\nn
&&
+w_q  {\rm tr}
\left[S_c^{[12]}\gamma^5 S_q^{[2]}\!\not\! p S_q^{[2]}\gamma^5 \right] 
{\rm tr}\left[ S_c^{[3]}\gamma^\mu S_q^{[13]}\gamma^\nu \right] 
\nn
&&
-w_c {\rm tr} \left[S_c^{[12]}\gamma^5 S_q^{[2]} \gamma^5 \right]
{\rm tr}\left[S_c^{[3]}\!\not\! p S_c^{[3]}
              \gamma^\mu S_q^{[13]}\gamma^\nu \right]
\nn
&&
+w_q  {\rm tr}\left[ S_c^{[12]}\gamma^5 S_q^{[2]} \gamma^5 \right]
{\rm tr}\left[S_c^{[3]} \gamma^\mu S_q^{[13]}\!\not\! p S_q^{[13]}
                        \gamma^\nu \right]
\Big\}
\label{eq:coupling}
\ena
where we have introduced the abbreviations 

\bea
S_c^{[12]} &=& S_c(k_1+k_2-w_c p),
\nn
S_c^{[3]} &=& S_c(k_3-w_c p),
\nn
S_q^{[2]} &=& S_q(k_2+w_q p),
\nn
S_q^{[13]}&=&  S_q(k_1+k_3+w_q p), 
\nn
&&\nn
K^2 &=&  \frac18 (k_1+2\,k_2)^2 + \frac18 (k_1+2\,k_3)^2 + \frac14 k_1^2\,. 
\nonumber
\ena
In the next section we shall describe how to evaluate the integral
(\ref{eq:coupling}).

\subsection{Infrared confinement}
\label{Sec:IR} 

In \cite{Branz:2009cd} we described how to integrate $n$-point one-loop 
diagrams and how to implement infrared confinement of quarks in this process. 
Since the present application involves also multi-loop diagrams
we need to extend our loop integration techniques to the case of arbitrary
number of loops. Let $n$, $\ell$ and  $m$ be the number
of the propagators, loops and vertices, respectively.
In Minkowski space the $\ell$-loop diagram will be represented as
\bea
&&
\Pi(p_1,...,p_m) = 
\nn
&=&
\int\!\! [d^4k]^\ell  
\prod\limits_{i_1=1}^{m} \,
\Phi_{i_1+n} \left( -K^2_{i_1+n}\right)
\prod\limits_{i_3=1}^n\, S_{i_3}(\tilde k_{i_3}+v_{i_3}),
\nn
&&\nn
&&
K^2_{i_1+n} =\sum_{i_2}(\tilde k^{(i_2)}_{i_1+n}+v^{(i_2)}_{i_1+n})^2
\label{eq:diag}
\ena
where the vectors $\tilde k_i$  are  linear combinations 
of the loop momenta $k_i$. The $v_i$ are  linear combinations 
of the external momenta $p_i$ to be specified in the following.
The strings of Dirac matrices appearing in the calculation need not concern 
us since they do not depend on the momenta. 
The external momenta $p_i$ are all chosen to be ingoing such that one has 
$\sum\limits_{i=1}^m p_i=0$. 
 
Using the Schwinger representation of the local quark propagator one has
\be
S(k) = (m+\not\! k)
\int\limits_0^\infty\! 
d\beta\,e^{-\beta\,(m^2-k^2)}\,.
\en
For the vertex functions one takes the Gaussian form. One has
\be
\label{eq:vert} 
\Phi_{i+n} \left( -K^2\right)\,
 =
\exp\left[\beta_{i+n}\,K^2\right] \qquad i=1,...,m\, ,
\en
where the parameters $\beta_{i+n}=s_{i}=1/\Lambda^2_{i}$ are 
related to the size parameters. The integrand in Eq.~(\ref{eq:diag})
has a Gaussian form with the exponential $kak+2kr+R$ where $a$ is
$\ell\times\ell$ matrix depending on the parameter $\beta_i$,
$r$ is the $\ell$-vector composed from the external momenta, and
$R$ is a quadratic form of the external momenta.
Tensor loop integrals are calculated with the help of the differential
representation 
\be
k_i^\mu e^{2kr} = \frac{1}{2}\frac{\partial}{\partial r_{i\,\mu}}e^{2kr},
\en 
We have written a FORM \cite{Vermaseren:2000nd} program that achieves 
the necessary commutations of the differential operators in a very efficient 
way.
 After doing the loop integrations one obtains
\be
\Pi =  \int\limits_0^\infty d^n \beta \, F(\beta_1,\ldots,\beta_n) \,,
\en
where $F$ stands for the whole structure of a given diagram. 
The set of Schwinger parameters $\beta_i$ can be turned into a simplex by 
introducing an additional $t$--integration via the identity 
\be 
1 = \int\limits_0^\infty dt \, \delta(t - \sum\limits_{i=1}^n \beta_i)
\en 
leading to 
\be
\hspace*{-0.2cm}
\Pi   = \int\limits_0^\infty\! dt t^{n-1}\!\! \int\limits_0^1\! d^n \alpha \, 
\delta\Big(1 - \sum\limits_{i=1}^n \alpha_i \Big) \, 
F(t\alpha_1,\ldots,t\alpha_n). 
\label{eq:loop_2} 
\en
There are altogether $n$ numerical integrations: $(n-1)$ $\alpha$--parameter
integrations and the integration over the scale parameter $t$. 
The very large $t$-region corresponds to the region where the singularities
of the diagram with its local quark propagators start appearing. 
However, as described in \cite{Branz:2009cd}, if one introduces 
an infrared cut-off on the upper limit of the t-integration, all 
singularities vanish because the integral is now convergent for any value
of the set of kinematic variables.
We cut off the upper integration at $1/\lambda^2$ and obtain
\be
\hspace*{-0.2cm}
  \Pi^c = \!\!  
\int\limits_0^{1/\lambda^2}\!\! dt t^{n-1}\!\! \int\limits_0^1\! d^n \alpha \, 
\delta\Big(1 - \sum\limits_{i=1}^n \alpha_i \Big) \, 
F(t\alpha_1,\ldots,t\alpha_n).
\en  
By introducing the infrared cut-off one has removed all potential thresholds 
in the quark loop diagram, i.e. the quarks are never on-shell and are thus
effectively confined. We take the cut-off parameter $\lambda$ to be the 
same in all physical processes.
The numerical evaluations have been done by a numerical program 
written in the FORTRAN code.

As a further illustration of the infrared confinement effect relevant to the 
applications in this paper 
we consider the case of a scalar one--loop two--point 
function. One has
\be
\hspace*{-0.2cm}
\Pi_2(p^2) =
\!\!\int\! \frac{d^4k_E}{\pi^2} 
\frac{e^{-sk_E^2}}{[m^2 + (k_E+\tfrac12 p_E)^2]
                  [m^2 + (k_E-\tfrac12 p_E)^2]} 
\nonumber
\en
where we have collected all the nonlocal 
Gaussian vertex form factors in the numerator factor $e^{-sk_E^2}$. Note that
the momenta $k_E$, $p_E$ are Euclidean momenta. Doing the 
loop integration one obtains 
\bea
\hspace*{-0.4cm}
\Pi_2(p^2) &=&\!\! \int\limits_0^\infty\!\! dt \frac{t}{(s+t)^2} 
\int\limits_0^1\!\! d\alpha \, 
\exp\Big[ - t z_{\,\rm loc}+ \frac{st}{s+t} z_1 \Big],
\nn
&&\nn
z_{\,\rm loc} &=& m^2 - \alpha(1-\alpha)p^2,
\qquad
z_1 =  \Big(\alpha - \frac{1}{2}\Big)^2 p^2.  
\ena 
The integral $\Pi_2(p^2)$ can be seen to have a branch point at $p^2=4m^2$
because $z_{\rm loc}$ is zero when $\alpha=1/2$. 
By introducing a  cut-off  on the $t$-integration one obtains  
\be 
\hspace*{-0.2cm}
\Pi^c_2(p^2) =\!\! \int\limits_0^{1/\lambda^2}\!\! dt \frac{t}{(s+t)^2} 
\int\limits_0^1\! d\alpha \, 
\exp\Big[ - t z_{\,\rm loc}+ \frac{st}{s+t} z_1\Big]\,.
\label{eq:conf}
\en 
The one-loop two-point function $\Pi_2^c(p^2)$ Eq.(\ref{eq:conf})can be seen
to have no branch point at $p^2=4m^2$. 

\section{The transitions \boldmath{$X\to \Jpsi+\rho(\omega)$} 
                     and \boldmath{$X\to D+\bar D^\ast$}  }

\begin{figure}[htbp]
\begin{tabular}{c}
\includegraphics[scale=0.4]{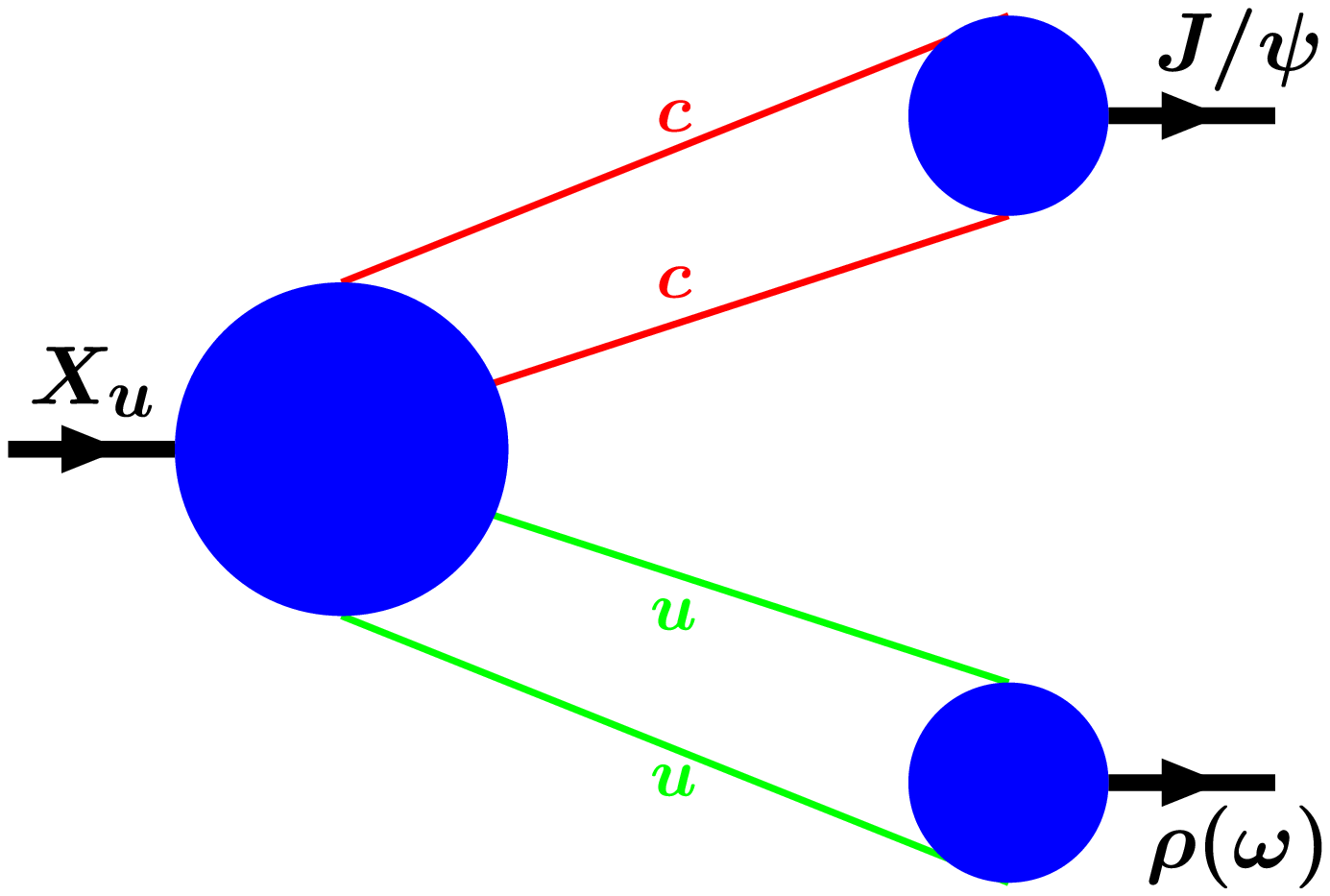} \\
\includegraphics[scale=0.4]{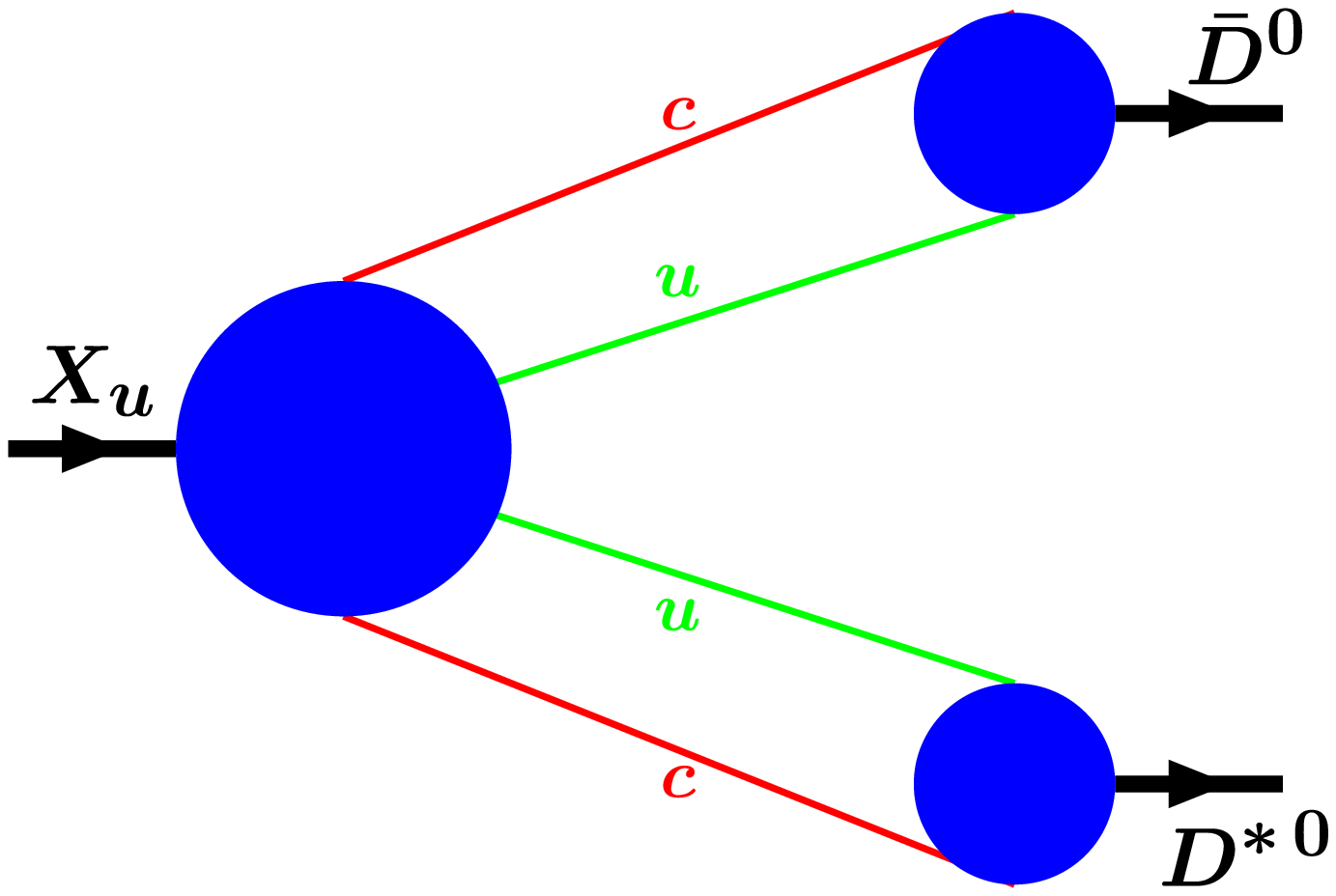}
\end{tabular}
\caption{
Feynman diagrams describing the decays
$X\to \Jpsi+\rho(\omega)$ and $X\to D+\bar D^\ast$.
}
\label{fig:decay}
\end{figure}

In this section we evaluate the matrix elements of the transitions
$X\to \Jpsi+\rho(\omega)$ and $X\to D+\bar D^\ast$. 
The relevant Feynman diagrams are shown in Fig.~\ref{fig:decay}.
Since the X(3872) is very close to the respective thresholds in both cases, 
{\it cif.} 
\bea
m_X-(m_\Jpsi+m_\rho) &=& -0.90\pm 0.41\,{\rm MeV},
\nn
m_X-(m_{D^0}+m_{D^{\ast\, 0}}) &=& -0.30\pm 0.34\, {\rm MeV}
\nonumber
\ena
the intermediate $\rho$, $\omega$ and $D^\ast$ mesons 
have to be treated as off-shell particles.

\bea
&&
M^{\mu\nu\rho}\Big(X_u(p,\mu)\to \Jpsi(q_1,\nu)+v^0(q_2,\rho)\Big)
\nn
&&\nn
&=& 6\,g_X\, g_{\Jpsi}\, g_{v^0}
\int\!\!\frac{d^4k_1}{(2\pi)^4i}\int\!\!\frac{d^4k_2}{(2\pi)^4i}
\bar\Phi_X\Big(-K_1^2\Big) \nn
&\times&
\Phi_{\Jpsi}\Big(-(k_1+\tfrac12 q_1)^2\Big)
\Phi_{v^0}\Big(-(k_2+\tfrac12 q_2)^2\Big) \nn
&\times&
{\rm tr}\Big[i\gamma^5 S_c(k_1)\gamma^\nu S_c(k_1+q_1)\gamma^\mu
S_u(k_2)\gamma^\rho S_u(k_2+q_2)\Big] \nn
&&\nn
&=&\,\,\,  \varepsilon^{q_1 q_2 \mu \nu} q_1^\rho\,M_{XJv}^{(1)}
   +\, \varepsilon^{q_1 q_2 \mu \nu} q_2^\rho\,M_{XJv}^{(2)}
\nn
&&
   +\, \varepsilon^{q_1 q_2 \mu \rho} q_2^\nu\,M_{XJv}^{(3)}
   + \varepsilon^{q_1 q_2 \nu \rho} q_1^\mu\,M_{XJv}^{(4)} 
\nn
&&
   +\, \varepsilon^{q_1 \mu \nu \rho}\,M_{XJv}^{(5)}
   + \varepsilon^{q_2 \mu \nu \rho}\,M_{XJv}^{(6)}
\nn
&&
   +\, \varepsilon^{q_1 q_2 \mu \rho} q_1^\nu\,M_{XJv}^{(7)}
   + \varepsilon^{q_1 q_2 \nu \rho} q_2^\mu\,M_{XJv}^{(8)} ,
\label{eq:XJV}\\
&&\nn
K_1^2 &=& \tfrac12(k_1 + \tfrac12 q_1)^2 + \tfrac12(k_2+\tfrac12 q_2)^2
                                       + \tfrac14(w_uq_1-w_cq_2)^2.
\nonumber
\ena
where $v^0=\rho,\omega$. In the case where the $X$ and $\Jpsi$ are on
mass-shell, i.e. $\epsilon_\mu(q_1^\mu+q_2^\mu)=0$ and
$\epsilon_\nu q_1^\nu=0$, the number of independent Lorentz structures
reduces to 6. Due to the obvious relations:

\bea
M(X_d\to\Jpsi+\rho)  &=& -\,M(X_u\to\Jpsi+\rho),  \nn
M(X_d\to\Jpsi+\omega)&=& \,\,\,M(X_u\to\Jpsi+\omega)
\nonumber
\ena 
one can express the decay amplitudes of the physical states defined by
 Eq.~(\ref{eq:mixing}) via  the decay amplitudes of the $X_u$ as
\bea  
M(X_l\to\Jpsi+\omega) &=&\,(\cos\theta + \sin\theta)\,M(X_u\to\Jpsi+\omega),\nn
M(X_h\to\Jpsi+\omega) &=&\,(\cos\theta - \sin\theta)\,M(X_u\to\Jpsi+\omega),\nn
M(X_l\to\Jpsi+\rho) &=&\,(\cos\theta - \sin\theta)\,M(X_u\to\Jpsi+\rho), \nn
M(X_h\to\Jpsi+\rho) &=&\!-(\cos\theta + \sin\theta)\,M(X_u\to\Jpsi+\rho).
\nonumber
\ena

Next we turn to the decay $X\to \bar D+D^\ast$. The generic matrix element
is written as 
\bea
&&
M^{\mu\nu}\Big(X_q(p,\mu)\to \bar D(q_1)+ D^{\ast}(q_2,\nu)\Big)
\nn
&=&
3\sqrt{2}\,g_X\, g_{D}\, g_{D^\ast} 
\int\!\!\frac{d^4k_1}{(2\pi)^4i}\int\!\!\frac{d^4k_2}{(2\pi)^4i}
\bar\Phi_X\Big(-K_2^2\Big) \nn
&\times&
\Phi_{D}\Big(-(k_1+w_c q_1)^2\Big)
\Phi_{D^\ast}\Big(-(k_2+w_c q_2)^2\Big) \nn
&\times&
{\rm tr}\Big[\gamma^5 S_c(k_1)\gamma^5 S_q(k_1+q_1)\gamma^\mu
S_c(k_2)\gamma^\nu S_q(k_2+q_2)\Big]
\nn
&+& (m_q\leftrightarrow  m_c, w_q\leftrightarrow w_c)
\nn
&&\nn
&=&\,\, g^{\mu\nu}\,M_{XDD^\ast}^{(1)} + q_1^\mu q_1^\nu \,M_{XDD^\ast}^{(2)}
  + q_1^\mu q_2^\nu \,M_{XDD^\ast}^{(3)}  
\nn
&+&\,\, q_2^\mu q_1^\nu \,M_{XDD^\ast}^{(4)} 
        + q_2^\mu q_2^\nu \,M_{XDD^\ast}^{(5)} \,.
\label{eq:XDDv}\\
&&\nn
K_2^2 &=& \tfrac18(k_1-k_2)^2 + \tfrac18(k_1-k_2+q_1-q_2)^2
\nn
&&
                             +\tfrac14(k_1+k_2+w_cp)^2.
\nonumber
\ena
%

Note that the decay of the X-meson into $D\bar D$ is forbidden.

\section{Numerical analysis}

\subsection{X-decays}

Using the matrix elements (\ref{eq:XJV}) for the decay 
$X\to\Jpsi+\rho(\omega)$ one can evaluate the decay widths
$X\to\Jpsi+2\pi(3\pi) $. We employ the narrow width approximation
which was extensively discussed in Ref.~\cite{Uhlemann:2008pm} 
and also used in Ref.~\cite{Faessler:2002ut}. One has 
\bea
&&
\frac{d\Gamma(X\to \Jpsi+n\pi)}{dq^2}  =
\frac{1}{8\,m^2_X\,\pi}\cdot \frac{1}{3}|M_{XJv}|^2 
\nn
&\times& \frac{\Gamma_{v^0}\,m_{v^0}}{\pi}
\frac{p^\ast(q^2)}{ (m^2_{v^0}-q^2)^2 + \Gamma_{v^0}^2\,m_{v^0}^2 }
{\cal B}(v^0\to n\pi), \label{eq:XJV_NWA}\\
&&\nn
&&
\frac{1}{3}|M_{XJv}|^2 =
\frac{1}{3}\sum_{\rm pol}|\varepsilon_X^\mu\,\varepsilon_{\Jpsi}^\nu \,
\varepsilon_{v^0}^\rho\, M_{\mu\nu\rho}|^2 .
\nonumber
\ena
Here $p^\ast(q^2)=\lambda^{1/2}(m^2_X,m^2_{\Jpsi},q^2)/2m_X$ is
the momentum of the $\Jpsi$ or the $(n \pi)$ system in the center-of-mass 
frame and 
$(n\,m_\pi)^2\le q^2\le (m_X-m_{\Jpsi})^2$ 
defines the kinematic region of the respective processes with $n=2$ for the 
$\rho$ meson and $n=3$ for the $\omega$ meson. 
The essential point which allows one to derive Eq.~(\ref{eq:XJV_NWA})
is the omission of polarization correlations. As was shown in
 Ref.~\cite{Uhlemann:2008pm}  they produce no effects if the
intermediate state is on-shell. Note that in our calculation we keep 
the $q^2-$dependence of the matrix elements as given by Eq.~(\ref{eq:XJV}). 
We use the masses and widths of the $\rho(\omega)$-mesons from \cite{PDG}
(all in MeV): $m_\rho=775.49$, $\Gamma_\rho==146.2$, ${\cal B}(\rho\to2\pi)=1$,
$m_\omega=782.65$, $\Gamma_\omega=8.49$, ${\cal B}(\omega\to 3\pi)=0.892$,

The adjustable parameters of our model are the constituent quark masses $m_q$,
the scale parameter $\lambda$ characterizing the infrared confinement
and the size parameters $\Lambda_M$. They were determined  by using
a least square fit to a number of physical observables, see 
\cite{Branz:2009cd}.
Below we display the numerical values for the parameters which are relevant
to the present paper.

\begin{widetext}
\be 
\begin{array}{cccc|cccccc|c} 
  m_{u/d} & m_s & m_c & \lambda & \Lambda_\pi & \Lambda_{\rho/\omega} &  
\Lambda_D  &  \Lambda_{D^\ast}  & \Lambda_{J/\psi} & \Lambda_{\eta_c} & \\ 
\hline 
\,\,\,0.217\,\,\, & \,\,\,0.360\,\,\, &\,\,\, 1.6 \,\,\, 
& \,\,\,0.181 \,\,\,&\,\,\,  
0.711\,\,\, &\,\,\, 0.295\,\,\, &\,\,\, 1.4\,\,\, &\,\,\, 2.3\,\,\, & 
\,\,\,3.3\,\,\, &\,\,\, 3.0\,\,\, &\,\,\,  {\rm GeV}\\
\end{array}
\label{eq:param}
\en 
\end{widetext}
Note that our fit values for the size parameters
of the pion and the $\rho$-meson are in qualitative agreement with those
found in the quark model based on the instanton vacuum 
Ref.~\cite{Dorokhov:2004ze}.

\begin{figure}[h]
\vspace*{0.5cm}
\includegraphics[scale=0.30]{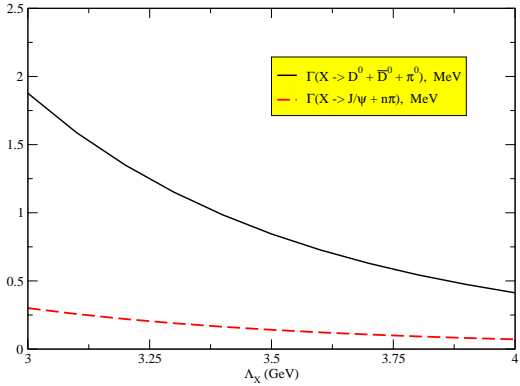}
\caption{The dependence of the decay widths 
$\Gamma(X_l\to \bar D^0 D^0 \pi^0)$
and $\Gamma(X\to \Jpsi+n\pi)$ on the size parameter $\Lambda_X$.}
\label{fig:width}
\end{figure}

There are two new free parameters: the mixing angle
$\theta$ in Eq.~(\ref{eq:mixing}) and the size parameter 
$\Lambda_X$. We have varied the parameter $\Lambda_X$
in a large interval and found that the ratio
\be
\frac{\Gamma(X_u\to \Jpsi+3\,\pi)} 
     {\Gamma(X_u\to \Jpsi+2\,\pi)} \approx 0.25
\en  
is very stable under variations of $\Lambda_X$. Hence, by using this result and
the central value of the experimental data given in Eq.~(\ref{eq:expt}),
one finds
\be
\frac{\Gamma(X_{l,h}\to \Jpsi+3\,\pi)}
     {\Gamma(X_{l,h}\to \Jpsi+2\,\pi)}  
\,\approx\,0.25\cdot \Big(\frac{1\pm\tan\theta}{1\mp\tan\theta}\Big)^2\approx 1
\en
which gives $\theta\approx \pm 18.4^{\rm o}$ for $X_l$~("+") and  $X_h$~("-"),
respectively. This is in agreement with the results obtained in both
\cite{Maiani:2004vq}: $\theta\approx \pm 20^{\rm o}$ and
\cite{Nielsen:2009uh}: $\theta\approx \pm 23.5^{\rm o}$.
The decay width is quite sensitive to the change of the size
parameter $\Lambda_X$. A natural choice is to take a value 
close to $\Lambda_{\Jpsi}$ and  $\Lambda_{\eta_c}$ 
which are both around 3 GeV.
We have varied the size parameter $\Lambda_X$ from  3 up to 4 GeV and 
display the dependence of the decay width in Fig.~\ref{fig:width}. 
One can see that
the decay width $\Gamma(X\to \Jpsi+n\,\pi)$ decreases  from  0.30 
up to 0.07 MeV, monotonously. This result is in accordance with 
the experimental bound  $\Gamma(X(3872))\le 2.3$~MeV and the result
obtained in \cite{Maiani:2004vq}: 1.6 MeV.

In a similar way we calculate the width of the decay 
$X\to D^0\bar D^0\pi^0 $ which was observed by the Belle Coll.
and reported in \cite{Gokhroo:2006bt}. 
Again using the narrow width approximation 
the differential rate reads
\bea
&&
\frac{d\Gamma(X_u\to \bar D^0 D^0 \pi^0)}{dq^2} =
\frac{1}{2 m^2_X \pi}\cdot \frac{1}{3}|M_{XDD^\ast}|^2 
\nn
&\times& \frac{\Gamma_{D^{\ast\, 0}}\,m_{D^{\ast\, 0}}}{\pi}
\frac{p^\ast(q^2)\,{\cal B}(D^{\ast\, 0}\to D^0\pi^0)  }
{ (m^2_{D^{\ast\, 0}}-q^2)^2 + \Gamma_{D^{\ast\, 0}}^2\,m_{D^{\ast\, 0}}^2 }, 
\label{eq:XDDv_NWA}\\
&&\nn
&&
\frac{1}{3}|M_{XDD^\ast}|^2  = 
\frac{1}{3}\sum_{\rm pol}
|\varepsilon_X^\mu\,\varepsilon_{D^{\ast\,0}}^\nu \,M_{\mu\nu}|^2 
\nonumber
\ena
where the matrix element $M_{\mu\nu}$ is defined by Eq.~(\ref{eq:XDDv})
and  $p^\ast(q^2)=\lambda^{1/2}(m^2_X,m^2_{D^0},q^2)/2m_X$ is
the momentum in the center-of-mass system. 
The kinematical region defined by 
$$(m_{D^0}+m_{\pi^0})^2\le q^2\le (m_X-m_{D^0})^2$$
is very narrow $3.99928\le q^2 \le 4.02672 $~GeV$^2$.  
Note that we have taken into account both channels with the intermediate 
$D^{\ast\, 0}$ and $\bar D^{\ast\, 0}$ mesons.
We use the masses and widths of the $D^{\ast\,+}$ and $D^{\ast\,0}$  mesons 
given in \cite{PDG} and \cite{Voloshin:2003nt}, \cite{Maiani:2007vr}
(all dimensional quantities in MeV): 
\begin{eqnarray*}
m_{D^{\ast\,+}} &=& 2010.27, \qquad \Gamma_{D^{\ast\,+}}=0.096,\\ 
m_{D^{\ast\,0}} &=& 2006.97, \qquad
{\cal B}({D^{\ast\,+}}\to D^+\pi^0) = 0.307,  \\ 
\Gamma_{D^{\ast\,0}}  &=& 0.070, \hspace{1.2cm}
{\cal B}({D^{\ast\,0}}\to D^0\pi^0)=0.619.
\end{eqnarray*}
Keeping in mind that
\begin{eqnarray*}
&&   
\Gamma(X_l\to \bar D^0 D^0 \pi^0)= 
\cos\theta^2 \Gamma(X_u\to \bar D^0 D^0 \pi^0),
\end{eqnarray*}
we have  varied  $\Lambda_X$ from  3 up to 4 GeV and found that
the decay width $\Gamma(X_l\to \bar D^0 D^0 \pi^0)$ decreases  from  1.88 
up to 0.41 MeV, monotonously. In Fig.~\ref{fig:width} we plot the dependence 
of the decay widths $\Gamma(X_l\to \bar D^0 D^0 \pi^0)$
and $\Gamma(X\to \Jpsi+n\pi)$ on the size parameter $\Lambda_X$. 
Using the results of \cite{PDG}
\bea
&&
10^5{\cal B}(B^\pm\to K^\pm X)\cdot{\cal B}(X\to \Jpsi\pi^+\pi^-)
= 0.95 \pm 0.19, 
\nn
&&
10^5{\cal B}(B^\pm\to K^\pm X)\cdot{\cal B}(X\to D^0\bar D^0 \pi^0)
= 10.0 \pm 4.0
\nonumber
\ena 
one calculates the rate ratio
\be
\frac{\Gamma(X\to D^0\bar D^0 \pi^0)}
     {\Gamma(X\to \Jpsi\pi^+\pi^-)}  = 10.5\pm 4.7
\label{eq:ratio}
\en
The theoretical value for this rate ratio depends on the size parameter
$\Lambda_{X}$ as Fig.~\ref{fig:width} shows. One has
\be
\frac{\Gamma(X\to D^0\bar D^0 \pi^0)}
     {\Gamma(X\to \Jpsi\pi^+\pi^-)}\Big|_{\rm theor}  = 6.0 \pm 0.2 \,,
\en
where the theoretical error reflects the $\Lambda_{X}$ dependence of the 
ratio. The ratio lies within the experimental
uncertaintities given by Eq.~(\ref{eq:ratio}).

\subsection{ $\Jpsi$-dissociation} 

\begin{figure}[htbp]
\includegraphics[scale=0.30]{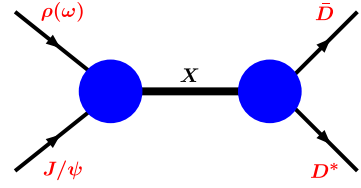} 
\caption{Diagram describing the X-resonance contribution
to the $\Jpsi$-dissociation process.}
\label{fig:X_diss}
\end{figure}

The last topic which we would like to discuss is the impact
of the intermediate X-resonance on the value of the $\Jpsi$-dissociation
cross section, see \cite{Barnes:2003vt}-\cite{J-diss-other}. 
The relevant s-channel diagram is shown in  Fig.~\ref{fig:X_diss}.  

\begin{figure}[h]
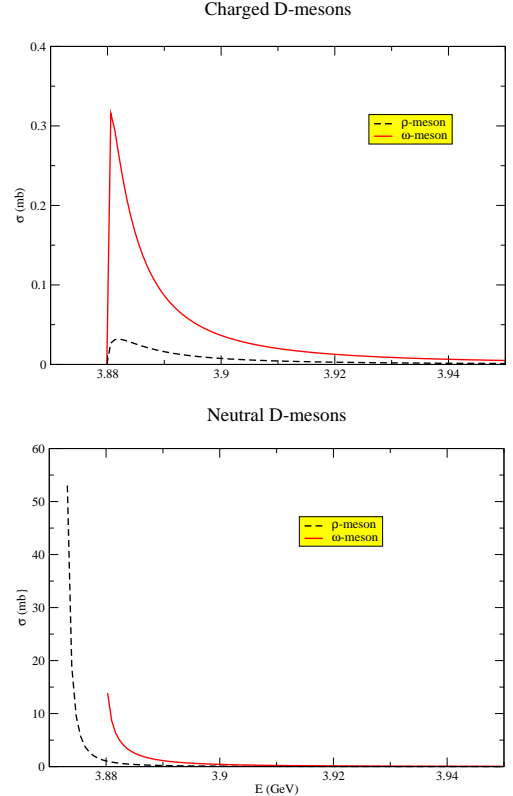

\begin{tabular}{cc}
\includegraphics[scale=0.28]{J_dis_charged.eps} \\
\includegraphics[scale=0.28]{J_dis_neutral.eps}
\end{tabular}
\caption{The cross sections of the processes
$\Jpsi+v^0\to X \to D + D^{\ast}$. Charged D-mesons--
upper panel, neutral D-mesons--lower panel.}
\label{fig:diss}
\end{figure}

We evaluate the $\Jpsi$-dissociation  cross sections
without using the narrow-width approximation.
One has
\bea
&&
\sigma(\Jpsi+v^0\to D+\bar D^\ast)+\sigma(\Jpsi+v^0\to \bar D+ D^\ast)
\nn
&=&
2\,(\cos\theta\mp\sin\theta)^2\,\sigma(\Jpsi+v^0\to X_u\to \bar D+ D^\ast),
\nn
&&\nn
&&
\sigma(\Jpsi+v^0\to X_u\to \bar D+ D^\ast)
\nn
&=&
\frac{1}{16\,\pi\,s}\frac{\lambda^{1/2}(s,m_D^2,m_{D^\ast}^2)}
                         {\lambda^{1/2}(s,m_{\Jpsi}^2,m_{v^0}^2)}
\frac{1}{9}\sum\limits_{\rm pol} \frac{|A|^2}{ (s-m^2_X)^2+\Gamma_X^2 m_X^2 },
\nn
&&\nn
&&
A=\varepsilon_{\Jpsi}^\nu\varepsilon_{v^0}^\rho\, M_{\mu\nu\rho}
\Big(-g^{\mu\alpha}+\frac{p^\mu p^\alpha}{m^2_X} \Big)
\varepsilon_{D^\ast}^\beta M_{\alpha\beta}
\label{eq:diss}
\ena
where $p=p_1+p_2=q_1+q_2$. $v^0=\rho$ (minus sign) or $\omega$
(plus sign). We also neglect isotopic spin breaking effects.
Note that $E=\sqrt{s}\ge m_{D^+}+m_{D^{\ast\,+}}$ for charged D-mesons in the final
states and $E\ge m_{\Jpsi}+m_{v^0}$ for neutral D-mesons in the final states. 
In the first case the cross section is zero at the threshold
of reaction $E= m_{D^+}+m_{D^{\ast\,+}}$. 
In the last case the cross section blows up at
$E = m_{\Jpsi}+m_{v^0}$ because the channel 
$\Jpsi+v^0 \to D^0 + \bar D^{\ast\, 0}$ is exothermic
and  the kinematical function
$\lambda^{1/2}(s,m_{\Jpsi}^2,m_{v^0}^2)$ in the denominator
is equal to zero at this point.
We take $\Gamma_X=1$~MeV in the Breit-Wigner propagator and
set $\Lambda_X=3.5$~GeV when calculating the matrix elements.
We plot the behavior of the relevant cross sections in Fig.~\ref{fig:diss}.
One can see that in the case of charged D-mesons (upper panel
in Fig.~\ref{fig:diss})  the maximum value of the cross section
is about 0.32~mb at  $E=3.88$~GeV. This result should be 
compared with the result of the cross section
$\sigma(\Jpsi+\pi\to D + \bar D^{\ast})\approx 0.9$~mb at 
$E=4.0$~GeV, see, \cite{Ivanov:2003ge} and the result of the cross section
$\sigma(\Jpsi+\rho\to D + \bar D^{\ast})\approx 2.9$~mb at 
$E=3.9$~GeV, see, \cite{Barnes:2003vt}. 
Thus the X-resonance gives a sizable contribution
to the $\Jpsi$-dissociation cross section. 
It would be interesting to do a complete analysis of the $\Jpsi$
dissociation cross section in view of our new results on the
s-channel contribution of the X(3872) tetraquark state. However,
this certainly is beyond the scope of the present  paper.

\begin{acknowledgments}

This work was supported by 
the DFG grant KO 1069/13-1,
the Heisenberg-Landau program,  
the Slovak aimed project at JINR and the grant VEGA No.2/0009/10.
M.A.I. also appreciates the partial support of the 
Russian Fund of Basic Research grant No. 10-02-00368-a. 
We thank P. Santorelli for his collaboration in the early
stages of this work. M.A.I. very much appreciates very interesting discussion
remarks and useful comments by 
A.E.~Dorokhov, 
G.V.~Efimov,
S.B.~Gerasimov,
N.I.~Kochelev,
D.V.~Shirkov and 
O.V.~Teryaev
during a "Hadron Physics" seminar held in
the Bogoliubov Laboratory of Theoretical Physics.

\end{acknowledgments}

\end{document}